\documentclass[referee]{aa}
\usepackage{graphicx}

\begin{document}

\thesaurus{07(07.19.1; 08.03.4; 08.16.2; 08.19.1)}

\newcommand{\mj}{${\rm M_J}$}
\newcommand{\ups}{$\upsilon$}

\title{Orbital migration and the frequency of giant planet formation}

\author{D. E. Trilling\inst{1}
\and J. I. Lunine\inst{2}
\and W. Benz\inst{3}}

\institute{University of Pennsylvania,
Department of Physics and Astronomy, David Rittenhouse Laboratory,
209 S.\ 33rd St., Philadelphia, PA 19120, USA (trilling@hep.upenn.edu)
\and
Lunar and Planetary Laboratory, University of Arizona,
Tucson, AZ 85721, USA
\and
Physikalisches Institut, Universitaet Bern,
Sidlerstrasse 5,
CH-3012 Bern,
Switzerland
}

\offprints{D. E. Trilling}

\date{Received / Accepted}

\authorrunning{Trilling et al.}
\titlerunning{Orbital migration and giant planet formation}
\maketitle

\begin{abstract}
We present a statistical
study of the post-formation migration
of giant planets
in a range of initial
disk conditions. For given initial conditions we model
the evolution of giant planet orbits under the influence
of disk, stellar, and mass loss torques. We determine the mass
and semi-major axis distribution of surviving planets after
disk dissipation, for various disk masses, lifetimes, viscosities,
and initial planet masses.
The majority of planets migrate too fast
and are destroyed via mass transfer onto the central
star. Most surviving planets
have relatively large orbital semi-major axes
of 
several AU or larger.
We conclude that the extrasolar planets observed to date, particularly
those with small semi-major axes,
represent only a small fraction ($\sim$25\%
to
33\%) of a larger
cohort of giant planets around solar-type stars, 
and many undetected giant planets must exist at large ($>$1-2~AU)
distances from their parent stars.
As sensitivity and completion of the observed
sample increases with time, this distant majority population
of giant planets should be revealed.
We find that the current distribution
of extrasolar giant planet masses implies that high
mass (more than 1-2~Jupiter masses) giant planet
formation must be relatively rare.
Finally, our simulations imply that
the efficiency of giant planet formation must be high:
at least 10\%
and perhaps as many as 80\%
of
solar-type stars possess giant planets during their pre-main sequence phase.
These predictions, including those for pre-main sequence stars, are
testable with the next generation of ground-
and space-based planet detection techniques. 
\end{abstract}


\section{Introduction \label{intro}}

Extrasolar giant planets (EGPs) have been detected
by the radial velocity method at orbital distances from
several AUs (e.g.,
47~UMa~c, Gl614b, $\epsilon$~Eri~b,
55~Cnc~d) to several hundredths
of an AU (51~Peg~b et al.)
from their central stars
(see
Mayor \& Queloz \cite{mq};
Butler \& Marcy \cite{bm};
Butler et al. \cite{butler97};
Noyes et al. \cite{noyes};
Cochran et al. \cite{cochran};
Santos et al. \cite{santos};
Vogt et al. \cite{vogt};
Hatzes et al. \cite{hatzes};
Butler et al. \cite{butler2001};
Santos et al. \cite{santos2001};
Fischer et al. \cite{fischer2002};
Marcy et al. \cite{marcy2002};
and many
others; see also the review by Marcy
et al. \cite{marcyppiv}).
Unless giant planets form in place within 1~AU of low mass stars --
unlikely in the context of published formation models (see,
for example, Guillot et al. \cite{guillot96}) --
the observed range of orbital semi-major axes implies that dramatic
orbital changes occur
after formation.
Previous work showed that in gaseous disks and even
subsequent particulate disks,  
giant planets can move from
formation distances of around 5~AU
to a wide range of final 
distances, a process known as
orbital migration
(Lin \& Papaloizou \cite{lp86};
Lin et al. \cite{linetal};
Takeuchi et al. \cite{takeuchi};
Ward \cite{ward97a,ward97b};
Trilling et al. \cite{trillingetal};
Murray et al. \cite{murray};
Bryden et al. \cite{bryden};
Kley \cite{kley99,kley2000};
Del Popolo et al. \cite{delp};
Tanaka et al. \cite{tanaka}).
The first direct determination
of the radius of an extrasolar giant planet via transit measurement
(Charbonneau et al. \cite{c00},
Henry et al. \cite{h00})
supports rapid inward migration of giant planets, since
the large planetary radius requires close proximity to the parent star
when the planet's internal entropy was much larger than
at present (Guillot et al. \cite{guillot96},
Burrows et al. \cite{burrad}), and formation in place is
generally considered implausible (see, e.g., Guillot et al.
\cite{guillot96}).
Early inward migration is commensurate with
the idea of migration caused by disk-planet interactions,
as disk lifetimes are not longer than $10^7$~years (e.g.,
Zuckerman et al. \cite{zuck}).

In this work, we address the following questions: (1)
How efficient is orbital migration? (2) What population
of planets survives the migration process?
and (3) How does this produced population compare to 
the observed EGP population?
We answer these questions by allowing 
a large
population of giant planets to evolve 
and migrate in circumstellar disks
with various initial conditions (one planet per disk)
and determining the final semi-major axis and mass
distributions.
We do not employ any stopping mechanisms, but instead
allow only those planets to survive whose migration
timescales are longer than their disks' lifetimes.
We compare the final semi-major axis and mass distributions
of the surviving model planets
to those of the observed EGPs.
We reproduce the fraction of planets
in small orbits, and predict the
frequency of and orbital distributions for
giant planets that are as yet unobservable.
Of those stars which do form giant planets, a few percent
should have giant
planets ultimately residing in small orbits,
and around one quarter of stars
which form planets retain planets in
orbits of semi-major axis several AU and beyond. Given the
current discovery statistics and their
uncertainties, our results imply
that giant planet formation is very efficient
around low-mass stars: we find that 10\% to 80\%
of young stars form planets,
with the uncertainty dominated by the initial
planetary mass distribution
(and with
some poorly-known
uncertainties associated with discovery 
statistics).
The presently detectable portion of the
giant planet population represents only
about 25\% to 33\%
of the total extant population of giant planets.

The model we present here is a simple one with a minimum
set of physical assumptions. By neglecting specific
stopping mechanisms, we provide results that represent
a baseline for planet formation statistics,
bereft of specific additional assumptions that any
given stopping mechanism would require. It is hoped
that the calculations and results presented 
here are transparent
enough that others can use them to explore
the particular effects that specific assumed
stopping mechanisms would have.
In short, we use the simplest possible
model in order to explore the consequences
of and for planet formation.

\section{Assumptions in the model}

In order to approach these problems in a simple
and clear way, we must make several assumptions
about the problem of planetary migration.
These assumptions are listed here and explained below:
(1) we use the simple impulse approximation for the Type~II
migration; (2) we employ no {\it ad hoc} mechanisms
to stop planetary migration;
(3) we assume that gas accretion onto planets migrating
in a gap is small, and neglect this accretion; and (4) we
neglect Type~I
migration.

\subsection{Impulse approximation}

In the migration model presented here, we use the 
impulse approximation described in
Lin \& Papaloizou (\cite{lp86}) to calculate
the torques between the planet and disk and determine
the planet's orbital motion.
We have previously shown (Trilling et al. \cite{trillingetal})
that the impulse approximation yields quantitatively
similar results to much more computationally intensive
methods of calculating orbital migration
(for example,
the WKB approximation of Takeuchi et al. \cite{takeuchi})
in the dissipation regime relevant to
this work.
Given our current knowledge of the migration
process, we prefer
to explore the consequences of the simplest possible
model rather than to add complexity which does not
guarantee more accurate results.

\subsection{Stopping mechanisms \label{stop}}

Halting
a giant planet's inward migration at small
semi-major axes
is even less well understood
than initiating and maintaining migration.
Various stopping mechanisms
have been proposed, including
tides
(Lin et al. \cite{linetal},
Trilling et al. \cite{trillingetal});
magnetic cavities
(Lin et al. \cite{linetal});
scattering by multiple planetesimals
(Weidenschilling \& Marzari \cite{weid});
resonant interactions by two
(or more) protoplanets
(Kley \cite{kley2000});
and
mass transfer from the planet onto the
star
(Trilling et al.
\cite{trillingetal}).
All stopping mechanisms have problems: Tides are insufficient 
because the magnitude of the tidal
torque is hardly ever enough to counterbalance
the inward torque;
magnetic cavities have not been demonstrated
to work in a quantitative way;
the presence of more than one planetesimal
or planet requires
a great deal of serendipity; and mass transfer
only works in a small percentage of cases.
Additionally, stopping mechanisms must
not only produce 51~Peg~b-like planets,
at small orbital distances, but must also
produce planets like HD108147b at 0.1~AU,
HD28185b at 1.0~AU, and
47~UMa~c at 3.8~AU
(Pepe et al. \cite{pepe};
Santos et al. \cite{santos2001};
Fischer et al. \cite{fischer2002}).
None of the above
mechanisms adequately halts planets at the
wide range of observed distances from the parent stars.

In this work, we presume that none of the above
mechanisms are effective in stopping more than
a small fraction of migrating planets, and instead consider
planets stopping for what is essentially a statistical
reason: 
the planet stops migrating when the disk
dissipates.
We show
that by considering a physically reasonable distribution
of initial conditions for planets
and disks,
our model 
produces a range of final orbital semi-major axes 
for extrasolar planets -- including 
small, 51~Peg~b-like separations --
without resorting
to any exotic stopping mechanisms.
It is not necessary to produce
many planets very close to their stars: the current detection
rate suggests that approximately 1\% of low-mass main sequence 
stars (F, G, K dwarfs) have extrasolar giant planets
in small orbits (Vogt et al. \cite{vogt}),
and extrasolar planets in all orbits exist around perhaps a few 
percent of all stars in the radial velocity surveys
(see, e.g., Marcy et al. \cite{marcyppiv}).
We note in passing that one
M-dwarf exhibits two companion objects of at least 
0.5~and
2~Jupiter masses (Marcy et al. \cite{marcy98};
Delfosse et al. \cite{delfosse};
Marcy et al. \cite{marcy2001a});
the statistics for M~dwarfs are thus very poor and indeed the 
environment for planet formation may be very different from that of
the more massive dwarfs. In what follows we confine ourselves to 
F~through K~stars and refer to these conveniently as low-mass stars.

\subsection{Planetary mass accretion during migration \label{acc}}

\begin{figure}
\resizebox{\hsize}{!}{\includegraphics[angle=-90]{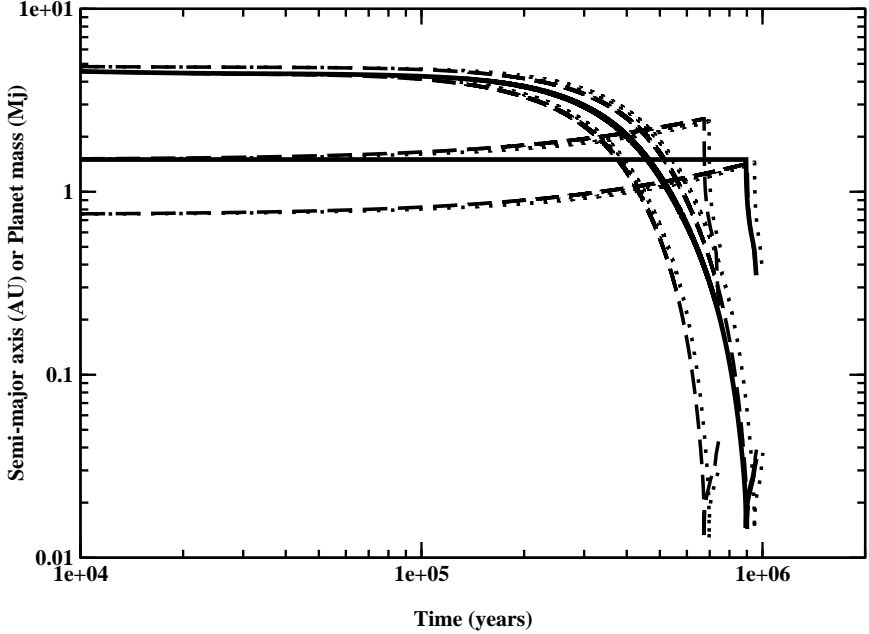}}
\caption{
Planetary semi-major axis (in AU) and planet mass (in \mj )
versus 
time for migrating planets.
Shown here are migrating planets' semi-major axes (generally
decreasing curves beginning at 5.2~AU)
and masses (generally flat or slightly
increasing curves beginning near 1~\mj )
for the following five cases:
the baseline case, with no mass accretion
during migration (solid black line);
two cases with linear mass accretion during
migration (dashed lines); and two cases
with exponential mass accretion during
migration (dotted lines).
For the no-accretion case, the initial
planet mass is 1.5~\mj .
For the accretion
cases, the initial planet masses are 0.75~\mj\
and 1.5~\mj ; the growth timescale is 
$10^6$~years.
The set of two semi-major axis curves above 
and to the right of 
the baseline case represent the two cases
with planets initially 0.75~\mj ;
the two semi-major axis curves below 
and to the left of the baseline curve
are planets with initial mass 1.5~\mj .
All planets lose mass through Roche lobe overflow
when their planetary radii are larger than the 
Roche radii.
This mass loss can be seen by a downturn in 
planet mass near the end of each run; mass
loss produces outward motion to conserve angular
momentum (thus the very late outward motion of
mass-loss planets).
Model runs were halted when planet mass became less
than 0.4~\mj .
This figure demonstrates that even allowing for 
mass accretion during migration,
the overall migration timescale
is not very different from that of the baseline (no
accretion) case. Because of this, for simplicity, we neglect 
mass accretion during migration
for the models in the statistical study.}
\label{newplot}
\end{figure}

Giant planets form gaps in their gaseous
circumstellar disks;
when this occurs,
$\dot{M_p}$, the planetary mass accretion rate, drops
substantially from its pre-gap value
as the local surface density is substantially reduced.
It is an outstanding question how much, if any,
material may flow across the gap from the disk
and be accreted onto the planet
(see Bryden et al. \cite{bryden};
Kley \cite{kley99};
Bryden et al. \cite{bryden2000};
Kley \cite{kley2000};
Lin et al. \cite{linppiv};
Kley et al. \cite{kley2001}).
The results shown in those
works suggest
that
accretion
timescales ($\tau_{acc}\sim M_p/\dot{M_p}$)
are between $3\times10^5$ and $3\times10^6$~years
for the viscosities, mass ratios, and other parameters
we use in our models.
We have adopted a characteristic accretion timescale
of $10^6$~years
and
performed several tests to determine the
effect
of including mass accretion in the migration models
(Figure~\ref{newplot}).
Note that the accretion timescale is comparable
to the migration timescale; we allow both linear
and exponential growth with this characteristic
timescale.
The results from these test cases show 
that when all disk parameters are held
constant, 
the range of variation
between the migration timescales
without mass accretion and 
timescales with mass accretion (all cases,
linear and exponential)
is around 25\% from the no-growth migration
timescale of $\sim10^6$~years
(Figure~\ref{newplot}).

Including mass accretion for a population
of migrating planets
has a small effect on the overall results, at best.
For more massive planets, which migrate
quite slowly anyway (see Sections~\ref{results}
and~\ref{asect}),
the effect of considering mass accretion
is very much smaller, or completely negligible.
Additionally, the primary parameters which
drive migration (diskmass, viscosity, disk lifetime)
are not known
to better than 25\%, and even the mass accretion
rate itself is a product of some underlying assumptions
about the flow of material near, around, and onto
a migrating planet.
Attention to the detail of mass
accretion during migration is not warranted
in the face
of these larger unknowns; instead,
we choose
to use a simpler model (ignoring post-gap
formation accretion) which may
neglect some details for the sake of clarity and understanding.
Lastly,
because
migration timescales are only affected by mass accretion
in a minor way, the overall
semi-major axis and final mass
distributions of the statistical study
(see below)
do not change significantly in the presence
or absence of mass accretion.
We therefore neglect mass accretion for the remainder
of the present work;
all subsequent discussions concerning the
statistical study implicitly refer to model
runs without mass accretion.

\subsection{Type I migration}

Giant planets, because of their large ratio of hydrogen and
helium relative to heavier elements, must form in gaseous protoplanetary
disks.
Regardless of whether a giant planet forms by
core accretion (e.g., Wuchterl et al. \cite{wuchterl})
or by direct collapse
(e.g., Boss \cite{boss97,boss98,boss2000,boss2001}),
when a planet grows to a certain mass ($\sim$10-30~$M_{Earth}$,
although there is some debate about what this critical
mass actually is),
a gap
is formed in the disk and spiral density waves travel
away from the planet
(see, for example, Goldreich \& Tremaine \cite{gt};
Lin \& Papaloizou \cite{lp86};
Ward \& Hourigan \cite{hourigan};
Lin \& Papaloizou \cite{lp93}).
Interactions of these
waves at the planet's Lindblad resonances
transfer angular momentum between the planet
and the disk; this angular momentum exchange in turn
produces a change in orbital distance of the planet.
Usually (but not always),
these interactions lead to a net loss of angular
momentum and a decrease in the planet's orbital
semi-major axis.
Such planet-disk interactions have been
described in detail
(see, for example,
Goldreich \& Tremaine \cite{gt};
Ward \& Hourigan \cite{hourigan};
Lin \& Papaloizou \cite{lp86,lp93};
Lin et al. \cite{linppiv};
Ward \& Hahn \cite{wardppiv}).
The type of migration executed with a planet in 
a gap is referred to as Type~II (see Ward \& Hahn \cite{wardppiv}).

So-called Type~I migration is executed by bodies which
are not massive enough to form gaps in their disks
(i.e., less than $\sim$10-30~$M_{Earth}$)
(Ward \cite{ward97a,ward97b}).
Ward has shown that Type~I migration may be very fast (less
than $10^6$~or even $10^5$~years) which is obviously 
detrimental to planet growth and survival: if migration 
timescales are so short for Earth mass objects,
it is unclear how any
planet could survive long enough to accrete enough gas mass
to open a gap
and
initiate the slower Type~II migration (which itself is 
perilous for planet survival).
Clearly, there is a problem with planet survival in the
face of Type~I migration. We do not intend to solve this problem
in this paper. Instead,
we neglect Type~I migration for the following reasons.

If Type~I migration were really as dangerously
efficient as has 
been proposed, then formation and survival of planets
(particular at distances greater than 0.1~AU)
would be nearly impossible, since the Type~I migration
timescales are shorter than any possible disk lifetime.
But, since we know that planets survived both at small
and intermediate
semi-major axes (many EGPs) and also relatively large
semi-major axes (Jupiter), Type~I migration cannot be
as universally destructive as proposed. 
For lack of a better quantification
of the Type~I effect, in the present work
we neglect
the effect of Type~I migration on the 
initial conditions used in our models.
Furthermore, all planets we consider in
this paper are too massive to undergo Type~I
migration.
Thus, 
we can
safely neglect Type~I effects 
in our calculations.
We reiterate that this does statement does not 
refute Type~I migration nor the importance of 
understanding and characterizing its timescales;
however, it is clear that for planets which have reached
$\geq$1~Jupiter mass
via any formation mechanism, Type~I migration
is no longer relevant.

\section{Planet migration and 
the statistical study of planet survival \label{stats}}

\subsection{The migration model and disk properties}

We have applied a model of disk-induced migration
(Trilling et al. \cite{trillingetal})
to a
set of planetary systems whose initial characteristics
are distributed over a representative range of
planet formation environments.
We use the same method and model as used in 
Trilling et al. (\cite{trillingetal}), which includes torques on
the planet due to disk-planet interactions, due 
to star-planet tidal torques, and due to planetary
mass loss.
Mass loss significantly prolongs planetary lifetimes for
only a very small population of model planets, but is included
here for completeness. Star-planet tidal torques affect
a larger population, but we shall see that, by far, the largest
effect is the magnitude of the disk-planet torque.
A planet's survival, which is enabled essentially entirely by
dissipation of the gaseous disk before
infall onto the star, depends on its
mass, and on the disk mass, disk lifetime, and disk
viscosity.

We have performed a statistical study of planet
fates in systems in which each of disk mass,
disk lifetime, and disk viscosity
is varied in turn. Initial planet masses are taken to be
integral values between 1 and 5 Jupiter masses 
(1~Jupiter mass~=~1~\mj\ =~$2\times10^{30}$~g), inclusive, evenly
distributed, with one planet per system.
Planetary accretion after gap formation
is neglected (see Section~\ref{acc}).
For the three varied parameters, we adopt
nominal values and vary each parameter
over a range spanning a factor of~30 larger
and smaller than the
the nominal value, with 200~model runs
evenly spaced
per decade in log space,
for a total of 600~model runs per
planet per varied parameter.
For example, in varying disk mass,
we adopt a nominal disk mass of 
0.02~$M_{\sun}$ (after Beckwith et al. \cite{beckppiv}),
and calculate
model cases between 0.02/30~and $0.02\times30~M_{\odot}$,
or 0.0007~and 0.6~$M_{\sun}$.
This range spans disk masses
from less than the
minimum mass disk for a system containing a Jupiter
to a disk so massive that direct binary star production
is more likely than giant planet formation
(see, for example, reviews by Beckwith \& Sargent \cite{bs}
and Beckwith et al. \cite{beckppiv}).
Over this entire range we test 600~model runs 
in evenly spaced intervals of 0.005~log units.
Similarly, we use a nominal disk lifetime and
alpha viscosity of $3\times10^6$~years and
$3\times10^{-3}$, respectively (after
Zuckerman et al. \cite{zuck}, Shakura \& Sunyaev \cite{shakura},
and Trilling et al. \cite{trillingetal})
and test parameter values
from disk lifetimes of $10^5$~to $9\times10^7$~years
and disk viscosities of $10^{-4}$~to $9\times10^{-2}$.
In all, we carried out 1800~model runs per planet mass
and 9000~model runs total;
each model run has one parameter varied
while the other two are given their nominal
values.
This very fine sampling of disk parameters
allows us to simulate a continuum of 
physical properties of planet-forming 
systems.
All planets' initial orbital semi-major axes
are 5.2~AU. Giant planets forming farther out would
spend more time migrating inward, all else being equal, and hence
would have a larger chance of survival (see below). 

The result of our computations
is a three-dimensional grid of 
models,
based on the nominal values and
spanning the range of
parameter space, for each initial planet mass.
To produce a
gaussian statistical population in which
the nominal value is most likely
and the extreme values are least
likely,
the significance of each model run 
was weighted by the gaussian
probability of the value of the varied
parameter according to the standard
gaussian formalism:

\begin{equation}
P(u)=\frac{1}{\sigma\sqrt{2\pi}}
e^{-\left \{ \frac{1}{2}\left(\frac{u-u_0}{\sigma}\right)^2 \right \} }
\end{equation}

\noindent where $u=\log~x$
and $u_0=\log~x_0$, $x$
is equal to the value of the varied
parameter, $x_0$ is the nominal value
for the parameter being varied, and
where
$\sigma$, which describes the 
width in log space of the gaussian parameter 
distribution we use, is equal to~0.5.
In other words, the half-width of the
gaussian physical parameter distribution is
a factor of
$\sqrt{10}$, or around~3.16;
thus, a parameter value which is 3.16~times
greater or less than the nominal value
has a probability one half of the 
probability that of the nominal value:
$P(u_h)=\frac{1}{2}P(u_0)$ for 
$u_h=u_0\pm0.5$ which is the same
as $x_h=x_0(3.16)^{\pm1}$. 

We use this gaussian weighting
in all results discussed in this paper.
Weighting is achieved by assigning
a significance to each surviving 
model planet as $\delta\times P(u)$
where $\delta$ is equal to 1~for 
planets which survive and 0~for planets
which do not.
Thus, in production
of histograms and discussing the 
ensemble results for the 9000~model
runs, we explicitly incorporate
the probability of the planet's
formation conditions and
circumstellar disk. Thus, planets
from extreme disks which survive
but whose initial conditions are extremely
unlikely are not overrepresented in
the statistical results.
Note that because probabilities are assigned
after all model runs are completed
and are simple functions of the nominal
values and a description of a gaussian
distribution,
other probability weightings
can be installed
to represent other
experiments.
Future work includes deriving both
$x_0$ and $\sigma$ more directly
from observations (see, e.g.,
Gullbring et al. \cite{gull};
Hartmann et al. \cite{hartmann})
and testing non-gaussian probability
distributions. 

Our models calculate orbital evolution
for $10^{10}$~model years, and 
planets are defined to have survived
when they have non-zero semi-major axes
after this time.
Very few planets
(much less than 1\%)
undergo significant orbital
evolution (due to star-planet
tides) after the disk dissipates.

\subsection{Results of the statistical study \label{results}}

\begin{figure}
\resizebox{\hsize}{!}{\includegraphics{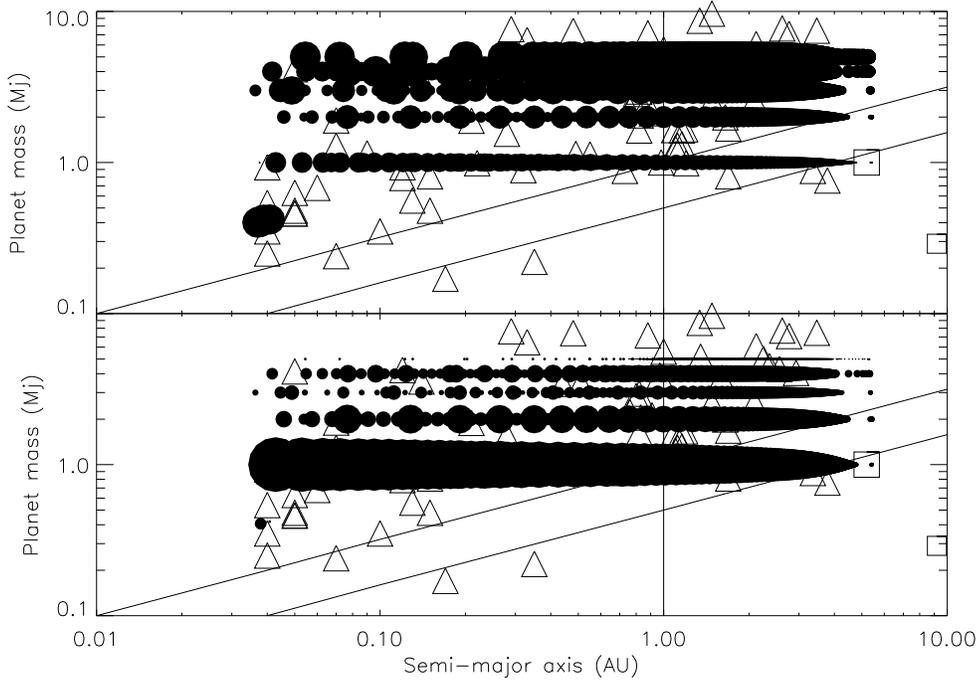}}
\caption{Planet mass versus heliocentric distance for
observed (triangles) and model
(filled circles) EGPs.
$M~\sin~i$ (minimum planet
mass) is used for all observed EGPs.
The open squares are Jupiter and Saturn.
Also shown are the 5- and 10-sigma
detection limits from the radial velocity
method of detecting extrasolar planets (lower
and upper diagonal lines). These
limits assume an RMS precision of 3~meters/second, which is
typical for the very high precision radial
velocity surveys
(Butler et al. \cite{butler96}).
The vertical solid line
at 1~AU corresponds to the completion
limit
of the radial velocity 
observations (G.\ Marcy, pers.\ comm.).
The two panels show results
for the flat (upper) and preferred
(lower) mass function (see text).
The radius of the point for each
model planet represents that
model's
probability (the same
arbitrary relative scaling
is used in both panels).
Beyond 1~AU, some EGPs have been detected,
but many others may also exist, undetected
to date.
Most observed EGPs are found at small semi-major axes because
these planets are easiest to detect.
The majority of model EGPs which survive
the formation and migration processes have large
semi-major axes (see text).
These two statements imply 
that a large population of EGPs
exists that has not yet been detected,
a prediction made by this work.
Observed EGP data is as of January, 2002.
Model masses are integral masses
1~\mj\ $\leq M_p\leq$ 5~\mj .}
\label{mfaf}
\end{figure}

The results from our statistical
study are shown in Figures~\ref{mfaf}
and~\ref{mfaf_alt}.
Of the 9000~initial planets, 
30\% survive. This survival rate
assumes a flat mass function (see below),
and
is normalized to the total probability
of all surviving planets; 
in this and subsequent discussions,
all results are post-weighting.
We find that
around 0.8\% of the initial planets have final
semi-major axes less than 0.1~AU.
Approximately 6.5\%
have final
semi-major axes between 0.1 and 1~AU.
The rest of the surviving planets
(23\% of the original planets) have
final distances from their stars greater
than 1~AU.
We find that 2.5\% of
all surviving planets
are found at less than 0.1~AU;
21\% are found between 0.1~and 1~AU;
and 76\% of surviving planets
have semi-major axes greater than 1~AU.
The majority
(70\%) of the population of initial
planets migrates too fast; these planets lose
their mass onto the central star (by Roche lobe
overflow), and do
not survive.
As noted in Section~\ref{stop}, other
stopping mechanisms at the inner edge of the disk may operate and would
increase somewhat the fraction of planets retained in orbits with
very small semi-major axes.
These mechanisms will have a small or negligible effect
on the overall
statistics, except for the possible
existence of a magnetic cavity, which effect on 
the migration has yet to be quantified.

\begin{figure}
\resizebox{\hsize}{!}{\includegraphics{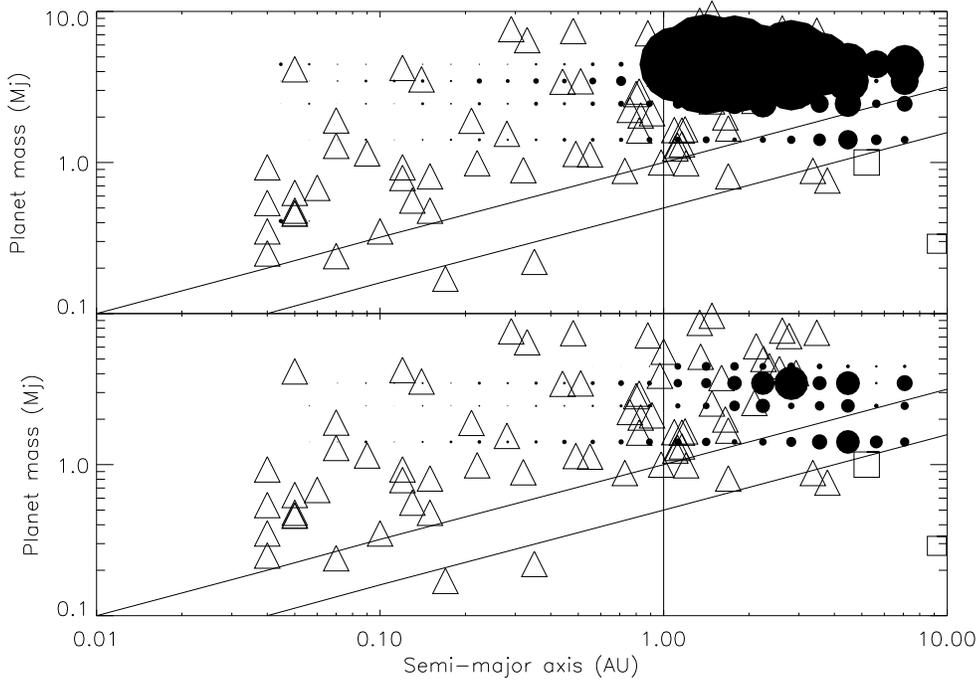}}
\caption{Same as in Figure~\protect{\ref{mfaf}}
but for the total overall probability of planet
survival in given semi-major axis and mass bins;
both semi-major axis and mass
bins are 0.1~log units
wide, centered on 0.05~log units.
The radii of the filled circles
again show the probability of model planets populating
that bin, but a different (arbitrary)
probability normalization is used
than for Figure~\protect{\ref{mfaf}}
to show circle size more clearly;
the same probability normalization is used
for both panels here.
As before,
the upper panel is shows results for
model planets with a flat initial mass distribution;
the lower panel is for results
with the preferred initial mass distribution.}
\label{mfaf_alt}
\end{figure}

The most likely fate of
a planet which forms near 5~AU is that it migrates inward
too fast and ultimately is lost onto the central star.
The vast majority of planets that do survive, though,
tend to reside at greater than 1~AU from
their central stars. Our statistical study
is robust because we have 
sampled planet migration over
a broad distribution of disk masses, viscosities, lifetimes
and initial planet masses: the variables that are key
to migration and its termination.
We have also carried out an initial  
study of the effects
of initial semi-major axis on planet migration
and survival. While
it is not easy or obvious to parameterize these
results,
it is clear that for identical initial disk conditions,
planets which form farther out migrate less rapidly.
This is because the surface density is lower at
larger semi-major axes, so that 
the torque between the planet and disk is less.
The effect on planet survival
is that more planets, of lower masses, survive (see below).
Therefore, if 5~AU is the minimum semi-major axis for
giant planet formation, then the results presented
here represent a minimum for planet survival;
conversely, if 5~AU corresponds to the largest
semi-major axis for planet formation, then 
the planet survival found here is the upper
limit.
We choose 5.2~AU as a nominal middle ground.
The fraction of planets which survive,
the ratio of planets at small semi-major axes compared to
at large semi-major axes, and the overall shape of the
surviving planet mass distribution are quite similar
even when somewhat different initial semi-major axes are considered
(see below).

\section{Final orbital semi-major axis distribution
\label{asect}}

\begin{figure}
\resizebox{\hsize}{!}{\includegraphics[angle=-90]{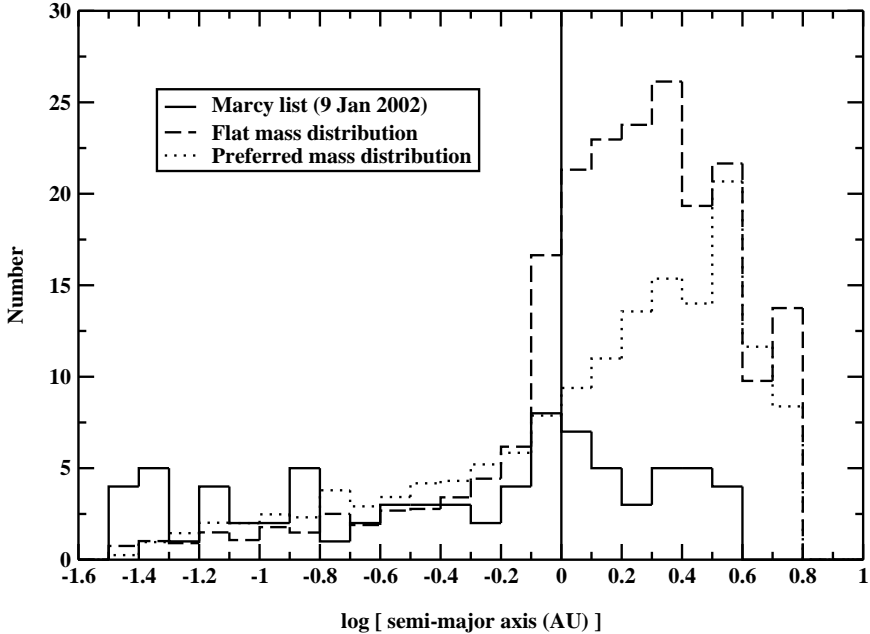}}
\caption{Histogram showing orbital semi-major axes
for observed (solid line)
and model (dashed and dotted) EGPs.
Each model histogram is normalized such
that the number of model planets with semi-major
axes less than 1~AU is equal to the number
of observed EGPs with $a<$~1~AU.
The vertical line corresponds to
1~AU, the completion limit for radial
velocity surveys.
Most model EGPs are to the right of the 
vertical line, i.e., at larger semi-major axes
than would have been detected thus far by the radial
velocity searches.
The dashed line is for a flat mass function;
the dotted line curve is for the preferred mass
function (see text).
Bins are 0.1~units in the log of semi-major axis.} 
\label{afhist}
\end{figure}

We have analyzed our statistical results in terms
of the final distribution of orbital semi-major axes
of the planets, in comparison to the observed
EGP distribution. The results
are shown in Figure~\ref{afhist}.
The observed EGP distribution
is relatively flat, that is, equal numbers of
planets throughout the range of 0.05~to
around 0.7~AU; there is the suggestion of 
a rise from 0.7~AU through just over 1~AU;
and a decrease at greater semi-major axes
than just over 1~AU (this decrease is 
due at least in part to observational bias).
(The
exception to this apparently flat
distribution is the
slight enhancement in planets at orbital distances
around 0.05~AU. In our previous work (Trilling et al. \cite{trillingetal}),
we described how tidal and mass loss torques can
cause a small ``piling-up'' of planets near the tidal
limit, which is typically around 0.05~AU.)
In contrast, the surviving planets in our statistical
study are predominantly at large ($>$~1~AU) 
semi-major axes, including a rise at
just under 1~AU to a peak at 2-4~AU.
The hint of an upturn in the observed distribution
at less than 1~AU echoes the pattern in the 
model distribution and may suggest the larger
population that we predict at greater than 1~AU.
The physical explanation for the
peak in model planets at separations greater than 1~AU
is that the time during
which a planet is migrating ($10^5$~to $10^6$~years)
can be relatively short compared
to both the lifetime of the disk ($\sim10^7$~years)
and the timescale required
to initiate migration (variable, but at least
$10^6$~years, or more) 
(Trilling et al. \cite{trillingetal}).
Therefore, the number of planets
which survive at small distances ($<$~1~AU) is small.
The majority of planets that survive
do so by 
creating very large gaps in their disks
and by 
requiring long timescales to excavate the
farthest regions of their 
large gaps, as follows.

More massive planets create larger gaps than
smaller planets, everything else being
equal
(Takeuchi et al. \cite{takeuchi};
Trilling et al. \cite{trillingetal}).
For the most massive planets, the length of time
it takes to fully form the gap can be quite long,
approaching or even surpassing the lifetime
of the disk, as such gaps are many AUs wide.
The gap is completely formed when
it reaches its equilibrium size
(see Takeuchi et al. \cite{takeuchi} and 
Trilling et al. \cite{trillingetal}).
During the stage in which a gap is growing
but has not yet reached its full equilibrium size,
there is no Type~I migration 
since the planet has already opened a gap;
there is also little 
Type~II migration
because 
most Lindblad resonances already fall within
the growing gap.
Thus, the planet is hardly migrating
during this time in which
the gap is still growing to its equilibrium size.
If the timescale for gap growth is longer than
the disk lifetime then the planet will
not migrate far.
Therefore we find that, overall, close-in
surviving model planets have smaller masses and
distant surviving models planets have larger masses;
this rough trend is seen in the observed
data as well (see below).
Note that relatively large mass planets (compared
to the disk mass) are
required, in general, for 
these arguments about large gap formation
to be applicable.

Since the vast majority of planets either migrate
too fast and are lost onto the star or remain out
near their formation location,
disk and planet parameters must
be ``just right'' to halt a planet
at small distances ($<$~1~AU).
The conclusion of this analysis is that
it is very difficult to make planets survive at
small distances from the parent star,
and that to do so requires a large population
of giant planets 
which still reside at large
distances from their parent star as well as
a population which has migrated too fast
and been lost onto the central star.
From our model results, we find that for
every planet found at small semi-major axes,
around 3~planets must exist at much larger
distances, depending on the initial mass
distribution (see below). 
The radial velocity technique is most sensitive to 
planets at small semi-major axes, because the 
magnitude of the stellar wobble is greatest for close-in 
planets. Additionally, since the highest precision
radial velocity surveys have relatively short
baselines of data, planets with longer periods
are only beginning to be identified
(e.g., 55~Cnc~d, with $\sim$15~year orbit
(Marcy et al. \cite{marcy2002})).
Therefore, the majority of the extant population of 
giant planets we predict
would, at present, 
be beyond the
detection capabilities of the radial velocity searches.
In other
words, a large population of Jupiter-like planets is as yet undetected.

\section{Implications}

\subsection{Giant planet mass distribution
\label{imf}}

We have also analyzed our surviving statistical
population for its distribution of final
planet mass, in comparison to the observed
EGP mass distribution. This comparison is
complicated by the fact that for most EGPs,
the known quantity is not planetary mass but
$M_p~\sin~i$.
We assume $\sin~i$~=~1,
for simplicity. We shall see in any case that it is not
the absolute mass that matters but the slope and
shape of the mass distribution curve.
The slope and shape
will be roughly unchanged between
$M_p~\sin~i$ and $M_p$ assuming a random distribution of
system inclinations relative to the Earth.
For a random distribution of inclinations, the
average of $\sin~i$ is $\pi$/4.
With increasing numbers of extrasolar 
planets known, the average $\sin~i$ for the
population gets closer to the value
for a random distribution. However, because
of the detection bias of the radial velocity
technique (biased towards detection of systems
with $i$~near 90~degrees), the average of $\sin~i$
for the detected systems is actually between $\pi$/4
and unity (Marcy et al. \cite{marcyppiv}).

\begin{figure}
\resizebox{\hsize}{!}{\includegraphics[angle=-90]{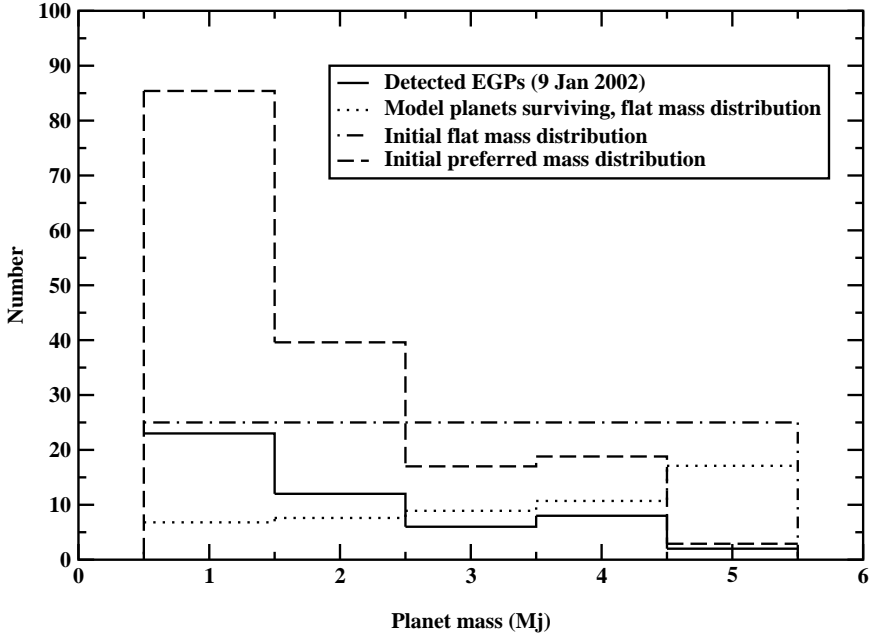}}
\caption{Histogram showing final planet mass distribution
for
observed EGPs (solid line).
Also shown are the
flat model
initial mass distribution (dot-dash line), 
the surviving population of model
planets produced from this flat
mass function (dotted line),
and the
preferred initial mass distribution (dashed line) required to
produce the observed mass function (see text).
For the observed EGPs, $M_p~\sin~i$ is
used.
The preferred
mass function 
is described in the text.
Mass bins are 1~\mj\
wide, centered on integral masses.
The model mass distributions are normalized
to the number of observed EGPs (solid line).}
\label{mfhist}
\end{figure}

The comparison among mass functions used
in this work is shown
in Figure~\ref{mfhist}.
By assumption, in our model we
have started with equal numbers of
planets with masses 1, 2, 3, 4, and 5~\mj\
(i.e., after accretion
has taken place), or a flat initial mass distribution.
As shown in Figures~\ref{mfaf},~\ref{mfaf_alt}, and~\ref{mfhist},
most of the surviving planets in the statistical
study are the more massive planets;
planets with the largest initial mass (5~\mj ) survive
preferentially.
The observed EGP mass distribution, however, is 
biased toward smaller mass planets (Figure~\ref{mfhist}).
Since the radial velocity
technique is most sensitive to larger masses, the
relative dearth of large mass planets discovered is
significant.
The present observed EGP population
can instead be obtained with 
an initial mass distribution
biased strongly toward smaller mass
giant planets. This
``preferred'' initial distribution is shown in Figure~\ref{mfhist}
as the dashed line.
The preferred mass function is derived from the
observed mass distribution and the ratio of initial model
planets to surviving model planets:

\[
\frac{{\rm \#~of~initial~preferred~mass~distribution~planets~with~M_i}}
{{\rm \#~of~observed~EGPs~with~M_i}}
=
\frac{{\rm \#~of~initial~flat~mass~distribution~planets~with~M_i}}
{{\rm \#~of~surviving~flat~mass~distribution~planets~with~M_i}}
\]

\noindent
where initial planet masses
are
${\rm M_i}$~=~1, 2, 3, 4, and 5~\mj\
and mass bins of $\pm$0.5~\mj\ are used.
The preferred initial mass distribution by definition
produces a final mass distribution identical
to that which is presently
observed.

Model planet masses and semi-major axes
produced using the preferred the
initial mass distribution are 
shown in Figures~\ref{mfaf} and~\ref{mfaf_alt}.
The final distributions of 
semi-major axis
and planet mass
produced using the preferred initial
mass distribution are shown
in
Figures~\ref{afhist}
and~\ref{mfhist}.
As with a flat initial mass distribution,
in order to produce the observed extrasolar planet
characteristics from 
the preferred initial mass distribution,
many planets with large
semi-major axes must be produced.
The preferred initial mass distribution
results in 0.2\% of all initial
planets surviving with separations less than
0.1~AU;
1.6\% of all initial planets survive
and have 0.1~AU~$<a<$~1~AU;
and 
3.8\% of all initial planets survive
and have $a>$~1~AU.
These percentages correspond to
4.4\%, 28\%, and 68\% of
all surviving planets to be found
in those three semi-major axis 
ranges, respectively.
Thus, the ratio of as-yet undetected planets
to detected planets should be around~2:1,
compared to a ratio of~3:1 obtained with
the flat mass distribution.
Thus, the total percentage of stars with planets
must be 3-4~times greater
than the percentage given by the
current observational results.
With the preferred initial mass
distribution, 5.6\% of
all initial planets survive at 
any semi-major axis; with the
flat initial mass distribution,
this total survival rate is 31\%.
With the preferred initial mass distribution,
fewer planets
survive than in the case of the flat initial mass
distribution
because the smaller mass planets that dominate the
preferred initial distribution preferentially
are
destroyed\footnote{
Ward (\cite{ward97a,ward97b}) concludes that Type~II migration
timescales are independent of planet mass. This
conclusion is valid in the limit that the planet's
mass is small compared to the disk mass, resulting in
a  
planet which is fully entrained in the viscous evolution
of the disk. In the case of a finite mass ratio between
planet and disk, however,
there is a feedback torque from the planet onto the disk,
and the planetary migration timescale is seen
to be a function of planetary
masses
(Lin \& Papaloizou \cite{lp86};
Trilling et al. \cite{trillingetal};
Bryden et al. \cite{bryden2000};
Nelson et al. \cite{nelson};
Kley \cite{kley2000};
D'Angelo et al. \cite{dang}).
In the regime
in which we are working, planetary migration timescales
are functions of planet masses in the sense that,
in general, smaller
planets migrate faster and more massive planets migrate
more slowly.
This is partially a function of gap size,
as described in Section~\ref{asect}.}.
Our results imply that there
must still be a substantial population of giant planets
at large semi-major axis that has not yet been
detected. The ratio of extant planets at large semi-major
axes to extant planets at small semi-major axes is less
when the preferred initial mass distribution is used (around 2~to~1)
compared to when the nominal (flat) initial mass distribution is
used (closer to 3~to~1).


\subsection{Planet-forming efficiency \label{eff}}

Planet forming efficiency, that is, the number
of stars that form giant planets, must be relatively
high. This is because many planets which form are
subsequently destroyed, and many planets which exist
have not yet been detected, as follows:

\[
\frac{1~{\rm planet~detected}}{100~{\rm stars}}
\times
\frac{3~{\rm to}~4~{\rm extant~planets}}{1~{\rm planet~detected}}
\times
\frac{20~{\rm planets~formed}}{1~{\rm to}~6~{\rm extant~planets}}
\sim
\frac{10-80~{\rm planets~formed}}{100~{\rm stars}}.
\]

\noindent The first term, the number of observed stars
which have close-in planets, is from Vogt et al. (\cite{vogt})
and Marcy et al. (\cite{marcyppiv}); the
second term, that 2~to 3~planets exist for each planet
which has been discovered so far, is from Sections~\ref{asect}
and~\ref{imf};
the third term,
the fraction of planets formed which survive,
is from Sections~\ref{results}
and~\ref{imf}.
Note that our model only
incorporates 1~planet per star,
so this means~10\%
to~80\% of F,~G, and~K
dwarf stars form (at least)
one giant planet.
This is a very large percentage, which bodes well for 
observing large numbers of giant planets around young stars (see below).
The primary source of uncertainty is
the third term, based on the 
unknown initial mass distribution.

Our model suggests a high efficiency of
planet formation.
Since planets more easily survive migration
in lower mass disks,
planet formation in such disks leads
to greater planet survival. Additionally,
the shape of the preferred initial mass
distribution 
suggests a natural cutoff towards larger
mass planets: larger mass planets may
be more difficult to form
(in agreement with
Lineweaver \& Grether (\cite{linew})).
In other words, giant planet 
formation must be a relatively efficient use of the disk gas: 
it cannot be a process
requiring primarily disks at the
upper end of the mass range we have considered,
but must also occur in disks with masses closer
to the minimum we have selected.
Our minimum mass disk is slightly less
than
the mass of Jupiter itself, and represents
an extremum. Exclusion of extreme diskmasses from the
cohort 
changes the overall required efficiency
very little because these extreme cases
have very low probabilities and
therefore do not contribute much to the
total survival probability.
Overall we require that the
small- or moderate-disk masses
participate in the formation of giant planets,
and these giant planets are then
overrepresented in the final distribution because of slower migration rates.

Finally, it is important to emphasize that 
high planet forming efficiency does not 
mean that most low mass stars currently
have Jupiter-like planets. Planetary migration
is very unforgiving: our statistical study shows
that around 75\% to almost 95\% of formed planets migrate too
fast and meet their demise at their central
stars. A large percentage (10\%~to~80\%) of stars and disks
must produce planets so that a few percent may survive close to
their central stars and be detected by radial
velocity searches. But there should then be 
a population of giant planets in intermediate (2-4~AU) and
larger orbits, around perhaps
2\%-3\% of F, G, and K dwarf stars.
This fraction agrees with the 
estimates of Zucker \& Mazeh (\cite{zm2}).
Giant planet formation
may be occurring around many young stars today,
and searches for indications of planet
formation around these stars should have a high
success rate.

\section{Discussion and some uncertainties \label{uncert}}

Our preferred initial mass distribution
and mechanism for producing close-in
giant planets requires a high frequency of planet formation
in disks, and a bias toward making giant planets with smaller
masses (in agreement
with Lineweaver \& Grether \cite{linew}).
The latter conclusion is robust so long as the distribution
of observed system inclinations as seen from Earth is 
roughly random; that
is, as long as the measured $M_p~\sin~i$ yields actual masses on
average less than a factor of two larger.
To date, five systems with known inclinations (HD209458,
55~Cnc, HD210277, $\rho$~CrB, and $\epsilon$~Eri)
all have $i\ga30$~degrees
and thus have mass factors less than~2
(Charbonneau et al. \cite{c00};
Henry et al. \cite{h00};
Trilling \& Brown \cite{55cnc};
Trilling et al. \cite{omnibus};
Greaves et al. \cite{greaves}).
Additionally, by tidal arguments, 
\ups\ And, HD75289, HD187123, 51 Peg, and
HD217107 are all constrained to have 
$i\ga30$~degrees
as well, although there are no direct observations as yet
to confirm this (Trilling \cite{egptides});
the Hipparcos data implies an inclination $\sim$25~degrees
for \ups ~And (Mazeh et al. \cite{mazeh}).
Additionally, from dynamical 
and astrometric arguments,
the multiple-planet systems Gl876,
HD168443,
and $\upsilon$~And likely
have inclinations that are far from 
face-on (Laughlin \& Chambers \cite{laugh2001};
Marcy et al. \cite{marcy2001b};
Chiang et al. \cite{chiang}).
Therefore, most systems for which inclinations
are known or suspected
appear to have mass factors less than
around~2.
Astrometric studies of EGP systems
show a range of results but
few certain inclinations, to date
(Zucker \& Mazeh \cite{zm2000};
Halbwachs et al. \cite{halb};
Gatewood et al. \cite{gatewood};
Han et al. \cite{han};
Pourbaix \cite{pour};
Pourbaix \& Arenou \cite{poura};
McGrath et al. \cite{mcgrath}).

Giant planet formation is as yet not well enough understood to 
permit prediction of 
a preferred initial mass distribution
as a function of disk mass.
Models in which core accretion precedes accumulation of gas require a 
substantial surface density of solids to permit accretion to occur
within the disk lifetime
(Lissauer \cite{lissauer}),
with the resulting gas accretion
occurring fairly rapidly once a threshold mass of solids is accumulated
(Wuchterl et al. \cite{wuchterl}). There
is thus no simple and direct link between the disk gas mass and the final
masses of the giant planets formed. The presence of Uranus and Neptune in
our own system is suggestive of a process that is sensitive to depletion of
the gas either due to small available amounts of gas beyond a certain 
distance from the primary, or to long core accretion times that stretch beyond
the disk lifetime. However, alternative models for the formation of Uranus
and Neptune exist that do not tie them directly to conditions in the 20-30
AU region of the gas disk (Thommes et al. \cite{thommes}). 

In sum, we do not as yet have a direct link between
the mass distribution
of giant planets and the mass of the disk. Thus our preferred
initial mass distribution
is a general prediction: whatever
mechanism forms giant planets tends to prefer the
smaller mass objects. This is at 
least crudely consistent with the inference that
the combined giant planet/brown dwarf mass function 
appears to show a minimum around approximately
10~\mj\ (Marcy \& Butler \cite{marbutl2000};
Zucker \& Mazeh \cite{zm1}),
suggestive of a planet-forming process biased to lower masses and a 
brown dwarf-forming process (direct collapse or disk instability) that is most
efficient at much higher masses.

In our study we did not consider minimum masses below 1~\mj .
The lack of data for objects significantly below $M_p~\sin~i$ of
0.5~\mj\ hampers our ability to extend the calculations,
and we must await further
and more sensitive data (e.g., SIM, see below) 
to establish the frequency of lower mass
planets.
However, the radial velocity surveys have
begun to 
detect planets with masses as small
as 0.12~\mj\ (HD49674b; Butler et al. \cite{butler2002}). 
This implies that orbital evolution takes place
for Saturn-mass and smaller planets.
Further observations will show the effects
of planet mass on orbital evolution.

Likewise, we have arbitrarily imposed planet formation at 5.2~AU. Formation
of giant planets out to double or even quadruple
that distance is possible, based on our
own solar system.
We have performed a small set of model
runs for planets with larger initial semi-major
axes, out to 10~AU. The overall behavior and survival
of these migrating planets 
is not qualitatively different from that 
of
the entire suite of 9000~models.
The general effect of larger initial semi-major
axes is an increase in the
survival rate (because migration times are longer --
see Papaloizou \& Larwood \cite{paplar}). This
would produce a
lower required planet-forming efficiency and an
initial mass distribution
which is biased 
toward slightly higher masses relative to the preferred
initial mass distribution,
since more
small mass planets survive.
Eventually, however,
a distance must be reached in all disks
at which accretion times become too long and
giant planet formation yields
first to ice giants (Uranus and Neptune) and then debris.
The current cohort of close-in giant
planets does not provide a constraint on the outer 
orbital radius at which giant planet
formation could occur.
A more complete sample of observed
EGPs at larger semi-major axes
is required.
It is possible that production of giant
planets is actually preferred at the position 
in the disk where water ice first condenses
out, because the surface density of ice 
cold-trapped there (and hence of total solids) is very high
(Stevenson \& Lunine \cite{stevlun}).
However, testing this hypothesis
must await a more complete sample of planetary systems.

Our model says nothing about the distribution of orbital eccentricities
of the known radial velocity companions. Many of the known planets with 
semi-major axes beyond the tidal circularization radius have
significant eccentricities. This suggests a process of orbital evolution 
commonly associated with the formation and earliest
evolution of giant planets, and yet one which failed
to act on our own four giant planets. 
The discussion of whether radial migration
in gaseous or particulate disks produces
eccentric orbits is an ongoing one (e.g.,
Ward \cite{ward97a,ward97b};
Bryden et al. \cite{bryden};
Kley \cite{kley2000};
Papaloizou et al. \cite{pap2001};
Murray et al. \cite{murray2002}).
The presence of high eccentricities suggests
that multiple giant planet formation
may be the norm, and that interactions 
among giant planets after disk migration
ends pump up the orbital eccentricities.
The stochastic nature of such interactions
precludes their inclusion in an analysis such as the one
we present. Further, until the
orbital inclinations can be determined, an important clue to the genesis of the
eccentricities is missing. 
Regardless of the mechanism(s) by which eccentricity is 
created, except for extreme cases,
our results regarding the semi-major axis distribution are hardly affected.
Our statistical results regarding planet formation
frequency hinge on the frequency of the
close-in giant planets versus those farther out.
Additional modest semi-major axis evolution
associated with eccentricity pumping does not
alter our results significantly. Dramatic
three-planet interactions that would leave
giant planets in vastly altered orbits
and greatly
affect our statistics are arguably rare. In summary, while we do not
calculate eccentricities induced by gas drag in our
one-dimensional model, the
results based on the statistics of 
semi-major axis distribution remain robust.

The migration behavior
and orbital evolution of 
planets in a multiple-giant planet system 
could differ from the results presented here.
In these cases, both the gas-planet interactions 
and the planet-planet interactions could cause
orbital evolution. Since we know that systems
with multiple giant planets exist (e.g., \ups\ And
(Butler et al. \cite{butler99}),
47~UMa (Fischer et al. \cite{fischer2002}),
HD168443 (Marcy et al. \cite{marcy2001b}),
55~Cnc (Marcy et al. \cite{marcy2002}),
and, to a lesser extent,
our own), 
quantitative studies of these systems are needed.
For some cases of multi-planet
systems, planet survival statistics
could be different from our results because of the additional
planet-planet interactions.
Additionally, if Type~II migration
should operate significantly more or 
less efficiently than we have calculated
here (either through the presence of another
planet or some other disk-planet interaction),
the overall survival and statistics we
have derived here could change substantially.

\section{Conclusions \label{final}}

We have presented a simple, baseline model of
planet migration. 
Our simple model allows us to study the statistical
behavior of planets migrating in disks without
complications introduced from relatively unconstrained
processes, like stopping mechanisms.
Our overall results can be useful as a baseline
statistical result of planet formation and survival.
We have shown that most giant planets, under nominal
initial conditions, migrate rapidly
relative to the disk lifetime and are destroyed
before the circumstellar disk dissipates. We have
found that
the observed extrasolar planets must represent
only the tip of the iceberg -- perhaps 25\% to 33\%
of the total extant population has been detected 
-- based on the required
distribution of initial parameters. Most planets
which have formed and survived reside outside the
current detection limit of the radial velocity
searches (a similar conclusion
was reached by Armitage et al. \cite{arm});
this prediction is supported with new
detections reported in Vogt et al. (\cite{vogt2002}).
We have shown that the planetary
initial mass distribution must be biased toward smaller
masses in order to produce the shape of the observed
EGP mass distribution, yielding our
preferred initial mass distribution
(see also Armitage et al. \cite{arm});
again, recent detections by
Vogt et al. (\cite{vogt2002})
support the conclusion that the lowest mass
planets are the most common.
Finally, because migration in high
mass disks destroys almost all giant planets
formed therein, low mass disks -- within a factor of~10
or so of the mass of Jupiter -- must be capable of
forming giant planets (essentially in
agreement with recent observational
results by Carpenter (\cite{carp})).
Hence we find that giant planet
formation must be a relatively efficient process
in disks. Because 
high mass planets
migrate more slowly than low mass planets
in a given disk,
overall there should be more massive planets
at intermediate and large semi-major axes
than found close-in.
The population of close-in planets,
in contrast,
should be dominated by smaller mass
planets.
Zucker \& Mazeh (\cite{zm2002})
have shown that this effect is, in
fact, observed and statistically significant.

Migrating giant planets may be detrimental to terrestrial
planet survival, if terrestrial planets form coevally with
giant planets.
Planets interior to a migrating giant planet would be
disrupted and lost from the system.
This of course assumes
that smaller planets do not migrate, 
although they too likely migrate, potentially
on even shorter timescales than giant 
planets (Ward \cite{ward97a,ward97b}).
If terrestrial planets form after the
dissipation of gas in the protoplanetary disk,
then disruption by a migrating giant planet
may be less of a risk (excepting giant planet
migration caused by planet-planetesimal interaction
(Murray et al. \cite{murray})).
Kortenkamp \& Wetherill (\cite{kort})
have considered the case of terrestrial
planet formation when Jupiter has both its current
and a larger heliocentric distance
of 6.2~AU. They have
found that accumulation of rocky bodies may
be easier with Jupiter at a larger
heliocentric distance (that is, pre-migration).
It is possible that formation of the terrestrial
planets in our Solar System may reveal clues about
giant planet migration in our planetary system;
certainly, these studies are also relevant to 
the formation of small, rocky planets in other
planetary systems.
In the context of large scale migrations, if 
terrestrial planet formation
requires that giant planets have not migrated
through the terrestrial zone around 1~AU, then only
around~20\% (flat initial mass distribution)
or~3\% (preferred)
of all planet-forming systems qualify 
(i.e., a few percent of all late-type stars; see sections~\ref{stats}
and \ref{eff}).
If, however, the
constraint is merely that there 
be no giant planet in the immediate vicinity of
the terrestrial planet zone 
at the onset of terrestrial planet
formation (perhaps $10^7$~years), then the vast
majority (93\% for flat initial
mass distribution, 99\% for preferred)
of 
planet-forming systems qualify.
This number corresponds to nearly
all planet-forming systems which
is around~10\% to~80\% of late-type stars.

Although radial velocity detections of more
distant giant planets will become possible
as the time baseline of observations increases,
astrometric techniques are more sensitive to 
giant planets in large orbits. 
SIM, the Space 
Interferometry Mission
(Danner \& Unwin \cite{sim}),
will do a thorough job of detecting
giant planets, Uranus-mass objects, and even smaller bodies
from small to large semi-major axes
($\sim$10~AU),
with maximum sensitivity achieved
for planets approximately 0.3~${\rm M_{Earth}}$ around
3-5~AU.
SIM's
ability to test
our predictions of the
preferred initial giant planet mass
distribution
will be limited 
largely by mission lifetime. However, SIM can also analyze
disks around young stars with high precision
(target resolution of 1~microarcsecond), perhaps mapping
out the signature of gaps created by migrating (or
non-migrating) giant planets and
giving us a rough snapshot of the time-dependent mass distribution
of planets during the migration phase itself.
(A 1~AU gap at 100~parsecs is 10~milliarcseconds.)
For giant 
planets in the largest orbits, i.e., 20-30 AU from their parent 
star, direct imaging techniques may be the only practical method
for detection since astrometric techniques would require baselines
of decades or more.  

Our results suggest that techniques to study planet
formation around
young stars -- radial velocity;
high resolution imaging of young stellar
systems; searches
for gaps such as with SIM and potentially
also the Space InfraRed Telescope
Facility (SIRTF);
searches for planetary outflows (Quillen
\& Trilling \cite{quillen}); or searches
for
other indirect evidence, like cometesimals scattered onto
stars (Quillen \& Holman \cite{qh}) --
should ultimately have a very high success rate. We anticipate
that observational data will show that planet
formation is taking place around
10\%~to~80\% of 
low mass pre-main sequence stars, and that
planet searches around main sequence stars will have a much
lower success rate. Planet formation is an ``easy
come, easy go'' business, with many planets created
and many planets destroyed, and with an
important
minority -- including
our own Jupiter -- surviving.

\begin{acknowledgements}

The authors thank 
Doug Lin, Peter Bodenheimer, Geoff
Bryden, Hubert Klahr, Geoff Marcy,
Fred Adams,
and Norm Murray for many useful discussions.
We thank an anonymous referee for 
useful suggestions.
D.E.T. acknowledges a NASA GSRP grant,
and NASA grants to Doug Lin, Peter
Bodenheimer, and Robert H. Brown.
J.I.L. acknowledges support from
the NASA Origins Program. 
W.B. acknowledges support from the 
Swiss National Science Foundation.
D.E.T. thanks E. Stein for
hosting him for useful work sessions,
and NASA's
IRTF which accommodated 
him for some of the writing of this paper.

\end{acknowledgements}

\end{document}